\documentclass[a4paper,prl,twocolumn]{revtex4}
\usepackage{bm}
\usepackage{graphicx}
\usepackage{epstopdf}

\newcommand{\be}{\begin{equation}}
\newcommand{\ee}{\end{equation}}
\newcommand{\bq}{\begin{eqnarray}}
\newcommand{\eq}{\end{eqnarray}}

\newcommand{\bc}{\begin{center}}
\newcommand{\ec}{\end{center}}

\newcommand {\eps} {\epsilon}

\newcommand{\G}{\Gamma}

\def\(({\left(}
\def\)){\right)}
\def\[[{\left[}
\def\]]{\right]}

\def\lan{\langle}
\def\ran{\rangle}

\begin{document}

\title{Properties of the perturbative expansion around the
  mode-coupling dynamical transition in glasses}

\author{Silvio Franz$^1$, Giorgio Parisi$^2$, Federico
  Ricci-Tersenghi$^2$ and Tommaso Rizzo$^2$}

\affiliation{$1$ Laboratoire de Physique Th\'eorique et Mod\`eles
  Statistiques, Universit\'e Paris-Sud 11, B\^at. 100, 91405 Orsay
  Cedex, France\\ $2$ Dipartimento di Fisica, Sapienza Universit\`a di
  Roma, INFN, Sezione di Roma I, Statistical Mechanics and Complexity
  Center (SMC) - INFM - CNR, P.le Aldo Moro 2, I-00185 Roma, Italy}

\begin{abstract}
In this letter we show how to perform a systematic perturbative
approach for the mode-coupling theory. The results coincide with those
obtained via the replica approach. The upper critical dimension turns
out to be always 8 and the correlations have a double pole in momentum
space in perturbations theory. Non-perturbative effects are found to
be very important. We suggest a possible framework to compute these
effects.
\end{abstract}


\maketitle

In the mean field theory of glasses there are two transitions, the
dynamical-replica transition, that corresponds to the mode-coupling
transition in the dynamics and happens at a temperature $T_c$, and a
thermodynamical transition (the Kauzmann transition) that happens at a
lower temperature $T_K$ \cite{CAV,MP}.  The dynamical-replica
transition can be identified by looking at equilibrium properties of
the system, i.e. its landscape; it corresponds to the formation of
local minima in the free energy and it is usually studied using
replicas \cite{KWT,KT,REP}.  The mode-coupling transition is defined
by the dynamical properties of the system. The two transitions are
related as far as in the mean field approximation the time needed to
escape from a local minimum of the free energy is infinite.
 
This picture is exact in many solvable models. However it should be
modified in the real world, where the mean field approximation is no
more exact. In this letter we do not address the fate of the
thermodynamical transition and we concentrate our attention on the
dynamical mode-coupling transition.

It is quite evident that in short range systems there is no dynamical
transition, exactly for the same reasons for which there are no
infinite lifetime metastable states. However if we neglect the
so-called ``activated'' process, the dynamical transition is
present. Moreover there is a very large amount of experimental and
numerical data that are well fitted by the predictions of the
mode-coupling theory, so that it is certain interesting to try to
understand which is the critical behaviour associated to the
mode-coupling transition.

In this letter we present a computation of the upper critical
dimensions and of the critical properties of the dynamical
mode-coupling transition: in the dynamics we consider only the mutual
dependence of quantities that do not depend explicitly on the time,
i.e. time has been eliminated parametrically as it happens in the
generalized fluctuation dissipation relations \cite{CUKU,FM}. We will
firstly present the results in the framework of the equilibrium
replica approach and we will later show how the same results hold for
the mode-coupling transition.

The critical behaviour at the dynamical transition stems from the
presence of dynamical heterogeneities \cite{DAS,SH,PL, srpsp,
  CPR,FP}. In the dynamics these heterogeneities are related to the
presence of correlated movements of cooperatively rearranging regions
\cite{AdGibbs} that have been observed both above and below the
critical region around $T_c$.  We are interested in getting precise
predictions on the properties of dynamical heterogeneities.

Let us start with the basic definitions. Given two configurations of
the coordinates (that we label with $\sigma$ and $\tau$), we indicate
with $q_{\sigma,\tau}(x)$ the similarity (overlap) of the two
configurations in the region of space around the point $x$ (many
different definitions are possible). Usually $q$ is equal to one for
identical configurations and it takes a small value for uncorrelated
configurations \cite{PL,CFP, CGV,PS}. For example we can take
$q_{\sigma,\tau}(x)$ to be one if a region around $x$ of size
$a$ \footnote{Typically the ratio of $a$ with the inter-particle
  distance is taken to be smaller that 1 and not far from the
  Lindemann constant.}  has the same particle content in the
configurations $\sigma$ and $\tau$; otherwise $q_{\sigma,\tau}(x)=0$,
if the particle content is different.

Let us consider the case where $\sigma$ is an equilibrium
configuration of the system and $\tau(t)$ is a configuration obtained
using some dynamics at time $t$ starting from the $\sigma$
configuration (i.e.\ $\tau(0)=\sigma$). If the dynamics is
non-deterministic, the configuration $\tau(t)$ will depend also on
some extra random variables $\eta$. For simplicity of notation we will
not indicate the dependence of $\tau(t)$ on $\eta$, unless we need it
in an explicit way. We can define
\be
C(t) = \overline{q_{\sigma}(x,t)}\, \ \mbox{where}
\ q_{\sigma}(x,t)\equiv \lan q_{\sigma,\tau(t)}(x)\ran \;.
\ee
Here the overline denotes the average over the Boltzmann distribution
of the initial configuration ($\sigma$) and the angular brackets the
average over $\eta$. $C(t)$ is the usual equal point (smeared over a
region of size $a$) density-density correlation. Approaching the
dynamical transition, $C(t)$ will decay slower and slower, and will
also develop a plateaux (as function of $\ln(t)$) at the value
$C_P$. This plateaux becomes infinitely long at the mode-coupling
temperature; below the mode-coupling temperature, neglecting
\emph{activated processes}, the correlation does not decay any more,
i.e.\ $\lim_{t\to \infty}C(t)\equiv C_\infty>C_P>0$.

For the study of dynamical heterogeneities it is usual to consider a
dynamical susceptibility $\chi_4(t)$ defined as
\be
 V \chi_4(t)= \overline{Q_{\sigma}(t)^2} - \overline{Q_{\sigma}(t)}^2=
 \overline{ (Q_\sigma(t) -C(t))^2}\ ,
\ee
where $V$ is the volume of the system and $Q_{\sigma}(t)$ is the space
integral of $q_{\sigma}(x,t)$. The quantity $\chi_4(t)$ is a measure
of the differences that are observed during the evolution from region
to region \cite{DAS,SH,PL, srpsp, CPR, FP}.  In a similar fashion we
can define the time dependent correlation function
\be
G_4(x-y,t)=\overline{q_{\sigma}(x,t)q_{\sigma}(y,t)} -C(t)^2\ .
\ee

The explicit time dependence is quite a complex problem that we do not
address in this letter; different dynamical exponents are involved and
they are known to be not universal (at least in the mean field
approximation). We will consider here only relations where the time is
not explicitly presents, as the dependence of $\chi_4$ on $C$, that
can be obtained by eliminating the time parametrically, e.g. by
plotting $\chi_4(t)$ versus $C(t)$ exactly in the same way as for the
fluctuation dissipation relations. A particular example of a time
independent quantity is $\chi_4^*$, that is defined as the maximum of
$\chi_4(t)$ (that happens at time $t^*$) (the corresponding
correlation will be denoted by $G_4^*(x)$).  Analogously we define a
$C$-dependent correlation function $G_4(x-y|C)$ as follows
\be
G_4(x-y|C(t))=G_4(x-y,t)\ .
\ee

We are interested in the universal properties of the previous
quantities when we approach the dynamical transition. We denote by
$\eps$ the distance in temperature from the dynamical transition and
we indicate from here on the dependence on $\eps$. We are interested
in the behaviour in the double scaling limit $\eps\to 0$ and $C\to
C_P$.  Within mean field theory, the susceptibility $\chi_4(C_P,
\eps)$ is divergent when $\eps\to 0$ from above \footnote{We
  implicitly assume that the maximum $\chi_4^*(\eps)$ happens in the
  critical region $C \approx C_P$.}. Given the obvious relations
$\chi_4^*(\eps)=\int G_4^*(x,\eps)\, dx$ the divergence of
$\chi_4^*(\eps)$ implies the existence of a divergent correlation
length $\xi(\eps)$. In the same way as in standard phase transitions,
we expect the following scaling laws
\be
\chi_4^*(\eps)\propto \eps^{-\gamma} \ ,\ \ 
\widetilde G_4^*(k,\eps)=\chi_4^*(\eps) \tilde g_4(k \xi(\eps))\ ,\ \ 
\xi(\eps) \propto \eps^{-\nu}\, ,\
\ee
where $\widetilde G_4^*(k,\eps)$ is the Fourier transform of
$G_4^*(x,\eps)$.

What is the replica counterpart of this behaviour? In the replica
approach \cite{REP} we consider two replicas $\sigma$ and $\tau$. We
denote by $H(\sigma)$ the original Hamiltonian, while the Hamiltonian
of the $\tau$ system is
\be
H(\tau)-h Q_{\sigma,\tau}\ .
\ee
The thermal averages are taken first with respect to the $\tau$
variables and later with respect to $\sigma$ variables.  The dynamical
phase transition is defined by the behaviour of $q_0(\eps)\equiv
\lim_{h\to 0^+}q(\eps,h)$. For $\eps>0$ we should have $q_0=q_{bulk}$,
i.e. a small value usually temperature independent; at $\eps=0$ we
should have $q_0(0)=C_P$ and for $\eps<0$ we should have
$q_0(\eps)>C_P$. This behaviour is present in mean field model where
metastable states do exist. It survives in the real world in the
approximation where metastable states can be observed. In other words
the two replicas system (with the replica $\sigma$ quenched respect to
the replica $\tau$) becomes critical at the same point where the
dynamics display the mode-coupling singularity.

One can sharpen the physical picture by introducing a potential $W(q)$
defined as follows \cite{REP,MP,PZ}. We consider an equilibrium
configuration $\sigma$.  We call $P_\sigma(q)$ the probability that
another configuration $\tau$ has an overlap $q_{\sigma,\tau}=q$. We
define
\be
W(q)=-\lim_{V\to\infty}{\ln(P_\sigma(q))\over V} \ .
\ee
With probability one when the volume $V$ goes to infinity, the
potential $W(q)$ does not depend on the reference configuration
$\sigma$. In other words $P_\sigma(q)\approx \exp (- V W(q))$.  By
construction $W(q_{bulk})=0$ and the vanishing of the potential $W(q)$
for more than one $q$ value is the distinctive characteristic of
replica symmetry breaking (this should happen below an eventual
thermodynamical glass transition).  In other words we are considering
an equilibrium configuration $\sigma$ and we define by $W(q)$ the
increase in the free energy density if we constrain an other
equilibrium configuration $\tau$ to stay at overlap $q$.

\begin{figure}[t!]
\includegraphics[width=\columnwidth]{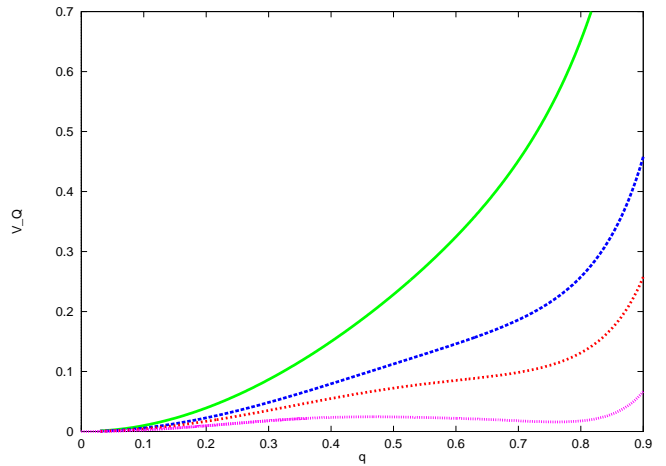}
\caption{ A schematic view of the potential $W(q)$ computed at various
  temperatures, decreasing from above to below.}
\label{W}
\end{figure}

The behaviour of the potential in the mean field approximation is
described in Fig.~\ref{W}. The dynamical transition is characterized
by the presence of an horizontal flex for the potential $W(q)$. Beyond
the mean field approximation the Maxwell construction should hold and
the non convex part of the potential disappear, but we will not
consider this effect.

A standard assumption is that the behaviour in the dynamics mirrors
the behaviour in the equilibrium properties of two replicas of the
system. This assumption is usually accepted and there are partial
proofs of its validity in perturbation theory (we will come later to
this point by showing how to complete these proofs). Let us try to
formulate this assumption in a sharp way.

In the same way as the time may be eliminated parametrically, also the
forcing field $h$ may be eliminated in favor of the expectation value
of the overlap (i.e. $q$) and all other physical variables can be
expressed as function of $q$; this is the usual Legendre
transformation of statistical mechanics. Our assumption is that in the
critical region ($\eps\to 0$ and $C\to
C_P$)the quantities
\be
\chi_4(\eps|q)\, ,\ \ \G_4(x-y,\eps|q) \, ,
\ee
computed  {\em in perturbation theory}  in the replica approach are the same as those computed in the
mode-coupling theory (as function of $C$). This relation holds only in
the region of $q$ where the forcing field $h \approx 0$.

Let us start the perturbative computation in the replica approach. The
computations in the mode-coupling approach will be rather similar. It
is quite clear from Fig.~\ref{W} that changing the temperature there
is a critical point. Our task it to compute the critical properties
using the standard renormalization group tools.  We want to determine
the critical exponents $\gamma$ and $\nu$ and the scaling function
$g_4(x/\xi)$, that according to the previous assumption are in the
same perturbative universality class in the mode-coupling theory and
in the replica approach. Let us discuss for the moment only the region
below the dynamical temperature, i.e. $\eps<0$, since this region is
very well defined in the perturbative expansion.

In high dimensions the correlation function is just given by the mean field
result \cite{CPR,BB0,BB1}
\be
\widetilde G_4(k,\eps)={1\over (|\eps|^{1/2}+k^2)^2}\;,
\label{doublep}
\ee
thus leading to the following mean field exponents
\be
\nu=\frac14  \qquad   \gamma=1\;.
\ee
As we will see, there is a crucial difference among the double pole
behaviour that we find to be valid in the general case and the single
pole behavior.  Notice that the double pole form in Eq.(\ref{doublep})
is at variance with the single pole behavior predicted in
\cite{BB0,BB1}, if no local conservation laws are present.  The
reasons for this discrepancy will be discussed later, when we confirm
the presence of a double pole by a dynamical analysis.

In order to compute the critical exponents of the dynamical transition
we will start from the perturbation theory around the mean field
theory. We will consider the most divergent diagrams near the critical
point and we will identify the universality class of the problem. In
this context we do not see that the critical point is in the
metastable phase because this would be a non-perturbative effect.

This computation can be done within the replica method of \cite{REP}
assuming a universal cubic effective replica action close to the
inflection point of $W(q)$ for $\eps\to 0$. For disordered systems
like Potts model, however, this model should be in the same
universality class of structural glasses as far the dynamical
transition is concerned.

We can define two correlations functions
\bq
G_0(x) &=&
\overline {\lan q_{\sigma,\tau}(x)\ran \lan q_{\sigma,\tau}(0)\ran}
-\overline {\lan q_{\sigma,\tau}(0)\ran}^2\;,\\ \nonumber
G_1(x) &=& \overline {\lan q_{\sigma,\tau}(x) q_{\sigma,\tau}(0)\ran}- 
\overline {\lan q_{\sigma,\tau}(x)\ran \lan q_{\sigma,\tau}(0)\ran}\;,
\eq
where the overline denotes the average over the $\sigma$'s and $\lan
\cdot \ran$ the average over the $\tau$'s.

In the case of a fluid with only one type of particles,
$q_{\sigma,\tau}(x)$ can be written as $\sigma(x)\tau(x)$ (where
$\sigma(x)$ and $\tau(x)$ are smeared densities around the point $x$)
and the previous equations become
\bq
G_0(x) &=&
\overline {\lan \tau(x)\ran \lan \tau(0)\ran\sigma(x)\sigma(0)} 
-\overline {\lan\tau(0)\ran \sigma(0)}^2\;,\\ \nonumber
G_1(x) &=& \overline {\lan \tau(x) \tau(0)\ran_c\; \sigma(x)\sigma(0)}\;.
\eq
One finds that near the critical point (neglecting loops)
\be
\widetilde G_0(k,\eps)={1\over (\eps^{1/2}+k^2)^2}\ , \ \ 
\widetilde G_1(k,\eps)={1\over \eps^{1/2}+k^2}\ .
\ee

In order to compute the loop corrections one must use a precise
replica setting. In the approach presented here, there is an annoying
asymmetry among the replica $\sigma$ and the replica $\tau$. In the
replica approach it corresponds to take one privileged replica
$\sigma$ and $n$ replicas of type $\tau$ and sending the value of $n$
to zero. It is known that a partially equivalent formalism consists in
using $m$ replicas that are constrained to be at overlap $q$ and take
the limit where $m$ goes to one. In this different formulation we have
a symmetry $Z_m$ in the limit $m \to 1$ and it may be more convenient
to use \cite{remi}. We checked that the two formulations are
equivalent near the stationary point and for simplicity we present the
analysis in this second formalism. A detailed discussion of this point
will be presented elsewhere \cite{FPRR}.

The diagrammatics is the same of an usual $\phi^3$ theory with two
differences \cite{CPR,DKT}: A) some of the propagators have a single
pole, others have double poles; B) the multiplicity of the diagrams
have to be computed in the limit where $m$ goes to one and some
diagrams give zero contribution in this limit.

One should do a careful analysis: at the end of a long analysis
\cite{FPRR} we find that many diagrams give zero contribution and the
dimension where the perturbative corrections are divergent is
$8$ \footnote{The value 8 for the the upper critical dimension was
  first suggested in \cite{CPR}. Then the value 6 was found in
  \cite{BB0}, while only in some particular cases (i.e. in presence of
  locally conserved quantities) the value was suggested to be equal to
  to 8 \cite{BB1}. }.  Moreover the diagrams are the same of those for
lattice animals, with the difference that here the effective coupling
constant is positive (for lattice animals is negative)
\cite{PAS1,PAS2,PH}.  The value 8 for the upper critical dimension may
provide an explanation for many of the anomalies found in \cite{SBBB}.

What happens below (and near) 8 dimensions?  The situation is quite
puzzling: the renormalization group pushes the coupling constant $g^2$
toward a large value and there is no perturbative fixed point that we
can analyze. Moreover the terms in the perturbation theory have all
positive sign and therefore it not easy to estimate the sum. This
result is not so disturbing.  Metastable states have finite life time,
the free-energy acquires an imaginary part that pushes the singularity
in the complex plane; in the same way the coupling constant takes an
imaginary part. Although the bare coupling is real, the fixed point
may correspond to an imaginary coupling constant.  One could argue
that asymptotically the exponent are like those of lattice animals for
the complex singularity.

However the previous conclusion may be to hasty. Using the same
arguments of \cite{PAS1,PAS2,PH} one finds that the sum of the leading
diagrams is related to the solution of the stochastic differential
equation governing the local fluctuations of the overlap
$\phi(x)=q(x)-q$ (being $q$ the space average of $q(x)$)
\be
 -\Delta \phi_\omega(x) + A+\eps \phi_\omega(x) +g \phi_\omega(x)^2=
 \omega(x) \label{STOA}
\ee
where $\omega(x)$ is a Gaussian short range noise, that is
$\overline{\omega(x) \omega(y)}=\delta(x-y)$. $A$ and $g$ are smooth
functions of the temperature and they are chosen in such a way to
implement the condition $\overline{\phi_\omega}=0 $. The two
propagators are given by the relations
\bq
G_0(x) &=& \overline{\phi_\omega(x) \phi_\omega(0)} \ , \nonumber \\ 
G_1(x) &=& \overline{\phi_\omega(x) \omega(0)} =
\overline{\left({1 \over-\Delta+\eps+2g\phi}\right)_{x,0}}\, ,
\eq
where the last equality follows from integration by part (and it is
correct only in case the solution to the stochastic differential
equation is unique).

Neglecting technicalities, the physics is quite clear. The choice of
the variables $\sigma$ (i.e. the initial conditions in the dynamics)
induce point dependent shift of the critical temperature and the
effects of these fluctuations is the dominant one.

One may wonder if there is a direct role of $ G_1(x)$ in the
dynamics. A suggestion is the following. Let us consider a theory
where the microscopic evolution equations for the particle have a
stochastic nature. In this case the overlap $q_{\sigma,\tau}(x)$ will
be a function of both the time $t$ and of the noise $\eta(x,t)$ and it
will be denoted by $q_{\sigma,\eta}(x,t)$.  We can define a different
dynamical susceptibility:
\be
V \chi_{22}(t)= \overline{\lan Q_{\sigma,\eta}(t)^2 \ran} -
\overline{\lan Q_{\sigma,\eta}(t) \ran^2}\;,
\ee
and the corresponding correlation $G_{22}(x,t)$, where the overline
denotes the average over initial conditions and angular brackets
average over $\eta$ \footnote{Notice the difference in the averaging
  procedure between the second addend in $\chi_{22}$ and the one in
  $\chi_4$.}. One could argue that in the region of time where $C(t)$
is near to (and above) the plateaux, $G_{22}(x,t)$ should behaves as
$G_1(x)$. On the other hand in the region where $C(t)$ is small, the
behaviour of $G_{22}(x,t)$ and $G_4(x,t)$ should be similar.

The whole analysis can be redone in the dynamical approach where one
takes care in an explicit way the dependence of the correlations on
the initial configuration (a fact that was neglected in \cite{BB0}).
The computations can be done in a neat way within the
Martin-Siggia-Rose (MSR) formalism of equilibrium dynamics using a
universal cubic dynamical action close to the mode-coupling
transition.  Detailed computations shows that taking care of the
correlations of the initial configuration with the evolving
configurations one recovers the same result of the replica
formalism. An explicit isomorphism of the two approaches can be shown
to be present: the relevant computations will be presented in
\cite{FPRR}.

Let us came back to the replica approach and let us consider in more
detail the stochastic differential equation (\ref{STOA}) that is
the resummation of the leading perturbative contributions. We are
interested to study it in a non-perturbative way.

It is well know that in perturbation theory the solution of the
stochastic differential equation is unique and that there an hidden
supersymmetry \cite{PAS1,PAS2,PH}. The supersymmetry relates the two
propagators and gives
\be
G_1(x)\propto {1\over x}{\partial G_0(x) \over \partial x}\ .
\ee
This supersymmetric relation is at the origin of the {\it dimensional
  reduction}: the critical exponents of lattice animals problem in
dimensions $D$ are the same of those of the Ising model near the
Lee-Yang singularity in dimension $D-2$.
 
However in presence of multiple solutions (as e.g. for the Random
Field Ising Model) everything becomes more complex (multiple solutions
cannot be seen in perturbation theory so that this problem does not
affect the perturbative analysis that we have presented above).
Supersymmetry and dimensional reduction are only valid if we average
over all the solution with a sign depending on the parity of the
corresponding Morse index: this weight is not the natural one from the
physical viewpoint \cite{PAS1,PAS2,PH}.
 
Which is the correct weight? If we stay within the replica formalism
it is useful to consider the free energy
\be
F[q]_\omega =\int dx \left( \frac12 {\partial^2 q(x) \over  \partial
   x^2}+ W(q(x)) - \omega(x) q(x) \right).
\ee
Equation (\ref{STOA}) can be written as $\delta F[q]_\omega / \delta
q(x) =0$. Now the natural choice would be to consider among the many
solutions the one that minimize $ F[q]_ \omega $. On the other hand
this choice is not natural in the dynamics where we would like to take
the solution $q_M(x)$ that maximize $q(x)$.  Indeed it can be proved
\cite{LMP} that there is a solution $q_M(x)$ such that $q_M(x)>q(x)$
for all $x$ and for any possible solutions of the stochastic
differential equation (this is a well know fact in the Random Field
Ising Model \cite{LMP}).  In both cases dimensional reduction is no
more valid and non-perturbative effects are present.

A neat formulation of the problem is the following: we introduce a
fictitious time $s$ and we write the following evolution equation:
\be
{\partial q(x,s) \over \partial s} = -\Delta q(x,s)
+ W'(q(x,s)) - \omega(x)
\label{FINAL}
\ee
with the boundary conditions at time $s=0$ given by $q(x,0)=1$.  In
the region below $T_c$, where the equation (\ref{STOA}) has many
solutions, the solution relevant for the dynamics is uniquely
identified as
\be
q^*_\omega(x)=\lim_{s \to \infty} q_\omega(x,s) \ .
\ee

We are near to the end of our journey. We still have to compute the
critical exponent and the critical behaviour of the correlation of
$q^*_h(x)$. Techniques introduced in \cite{PP} could be used to
achieve this goal.

We remark that the fictitious time $s$ in the equation (\ref{FINAL})
should not be identified with the real time: the dependence of $q(s)$
if we solve equation (\ref{FINAL}) is quite different from the
behaviour of $q(t)$ in mode-coupling theory and there should be no
confusion among the two variables. However one may make {\em en
  passant} the conjecture that, if we consider only reparametrization
invariant quantities, the mode-coupling equations and
eq. (\ref{FINAL}) stay in the same universality class: in other words,
if in the mode-coupling theory we eliminate the real time $t$ in
favour of $q$ and we write everything as function of $q$ and in
eq. (\ref{FINAL}) we eliminate the fictitious time $s$ in favour of
$q$, the dependence of the physical quantities on $q$ should be the
same near the critical point for the models. This conjecture is
trivially valid if there is only one solution of equation
(\ref{STOA}). The point is to understand its correctness beyond
perturbation theory
  
This conjecture may be generalized by introducing a more general
evolution equation that can be used to compute the properties of
reparametrization invariant quantities when activated processes are
present below $T_c$:
\be
{\partial q(x,s) \over \partial s}=-\Delta q(x,s)+
W'(q(x,s))-\omega(x)+\eta(x,s) \label{ACTIVATED}
\ee
where $\eta(x,s)$ is a thermal noise. The discussions of the
consequences of these conjectures cannot be done here, however it has
not escaped to how attention that eq. (\ref{ACTIVATED}) can be used to
explain the experimental results of \cite{EXP}.

Summarizing, we have clarified the predictions and the limitations of
the perturbative expansions for the critical properties of glasses,
finding an explicit mapping among the replica formalism and the
mode-coupling approach in the framework of the MSR approach to the
dynamics. We have also conjectured a mapping with a differential
stochastic equations that should be valid beyond perturbation theory,
whose interesting consequences should be studied in details.

\end{document}